\documentclass[nofootinbib,aps,amssymb,floatfix,superscriptaddress,preprintnumbers]{revtex4}

\usepackage{graphicx}
\usepackage{bm}
\usepackage{amsmath}
\usepackage{amssymb}
\usepackage{amsfonts}
\usepackage{float}
\usepackage{hyperref}
\usepackage{dsfont}  
\usepackage{slashed}  
\usepackage{booktabs}
\usepackage{multirow}
\usepackage{subfigure}
\usepackage[sort&compress]{natbib}
\usepackage{xcolor}
\usepackage{ulem}
\usepackage[percent]{overpic}

\newcommand{\be}{\begin{equation}}  
\newcommand{\ee}{\end{equation}}  
\newcommand{\beq}{\begin{eqnarray}} 
\newcommand{\eeq}{\end{eqnarray}}

\newcommand{\Slash}[1]{{\ooalign{\hfil/\hfil\crcr$#1$}}}
\newcommand{\nn}{\nonumber \\}

\newcounter{RSQ}

\begin{document}

\title{Single spin asymmetry in $e+p\to e'+B^\uparrow+X$ }

\author{Yoshitaka Hatta  }
\affiliation{Physics Department, Brookhaven National Laboratory, Upton, NY 11973, USA}
\affiliation{RIKEN BNL Research Center, Brookhaven National Laboratory, Upton, NY 11973, USA}

\author{Oleg V. Teryaev}
\affiliation{ Bogoliubov Laboratory of Theoretical Physics,
Joint Institute for Nuclear Research, 141980 Dubna, Russia}

\begin{abstract}

We study an exotic type of single spin asymmetry in unpolarized electron-proton scattering, in which the outgoing electron momentum exhibits a left-right asymmetry relative to the transverse spin of the leading baryon $B$ in the target fragmentation region.
We lay out two theoretical frameworks for describing this effect: The twist-three fracture function at high-$Q^2$ and the spin-dependent odderon in the high energy limit. 

\end{abstract}

\maketitle

\section{Introduction}

Transverse Single Spin Asymmetries (SSAs) are probably the most subtle polarization effects probing the finest details of the strong interaction. Their remarkable property is that they are `naively' odd under the time reversal operation T \cite{Christ:1966zz,DeRujula:1971nnp}. Distinct from  genuine 
T-violation (CP-violation) at the fundamental level, naive T-odd effects are consistent with the symmetries of QCD and caused by the appearance of imaginary parts  (phases) in the underlying scattering amplitudes. In the framework of collinear QCD factorization, such phases emerge from short or large distances or their overlap \cite{Efremov:1981sh,Efremov:1984ip,Qiu:1991pp,Qiu:1991wg}. Due to its twist-three nature, the calculation of SSAs quickly becomes very cumbersome beyond the leading order. Currently, next-to-leading order results are available for selected observables   \cite{Vogelsang:2009pj,Kang:2012ns,Yoshida:2016tfh,Gamberg:2018fwy,Benic:2024fvk,Rein:2025pwu,Zhang:2025fvt}.

The canonical definition of SSA is that, when an unpolarized projectile  collides with a transversely polarized target, the angular distribution of produced particles exhibits a left-right asymmetry with respect to the target spin direction. However, the polarized particle needs not be in the initial state. Even in unpolarized collisions,  one can consider asymmetries with respect to the spin of one of the final state particles.  In fact, such a phenomenon was predicted by Christ and Lee in 1966 \cite{Christ:1966zz}. They considered the exclusive production of a nucleon excited state $e(\ell)+p\to e'(\ell')+N^*$ and looked for an asymmetry of the form $(\vec{\ell}\times \vec{\ell}')\cdot \vec{s}_{N^*}$ where $\vec{s}_{N^*}$ is the spin vector of $N^*$. Their original motivation was to test  fundamental T-violation, but they realized  that there are backgrounds from the strong interaction that need to be subtracted. In practice, the final spin vector $\vec{s}_{N*}$ must be inferred from the decay  $N^*\to N+\pi$ which induces an absorptive part in the amplitude. This  interferes with the non-resonant continuum and results in the same angular correlation nowadays understood as a naive T-odd effect. Therefore, the Christ-Lee proposal does not provide a background-free  test of true T-violation and seems to have been largely forgotten. 
Note, however, that the mentioned  angular modulation may be considered as a correlation between final nucleon polarization and the  tensor polarization of the virtual photon. A  similar correlation between the  proton polarization and the tensor polarization of the deuteron is considered as a probe for a specific T-violation effect in low energy proton-deuteron  scattering \cite{Uzikov:2015aua,Nikolaev:2020wsj,Vergeles:2022mqu}.

As the present authors are  primarily interested in naive T-odd effects, not as backgrounds to new physics but as a fundamental topic of QCD spin physics, we propose to generalize the idea of Christ and Lee  \cite{Christ:1966zz}   to a new class of SSA observables associated with final state polarization. 
Specifically, we study the  asymmetry   $\sin (\phi_{\ell'}-\phi_{s_B})$  in unpolarized DIS $e+p\to e'+B^\uparrow+X$ where $B$ is a transversely polarized  leading baryon in the target fragmentation region such as the proton, neutron or $\Lambda$. We first present a general formulation of the problem in terms of the nonperturbative `T-odd fracture function'  (TOFF)   
\cite{Teryaev:2002wf,Anselmino:2011ss,Anselmino:2011bb,Chen:2021vby,Chen:2023wsi}. We then provide two specific mechanisms in DIS that contribute to the TOFF. One is the collinear twist-three fracture functions similar to the twist three distribution/fragmentation functions in `ordinary' SSA \cite{Efremov:1981sh,Efremov:1984ip,Qiu:1991pp,Qiu:1991wg}.  The other is the spin-dependent odderon \cite{Zhou:2013gsa,Boer:2015pni,Dong:2018wsp,Yao:2018vcg,Boussarie:2019vmk,Hagiwara:2020mqb,Kovchegov:2021iyc,Benic:2024fbf,Mantysaari:2025mht,Bhattacharya:2025bqa} in the high energy limit of DIS.

Crucial to our work is the assumption that the spin of the outgoing baryon $B$ can be experimentally measured. Although this is highly nontrivial, 
such `recoil polarimetry' has been successfully implemented at Jefferson laboratory \cite{JeffersonLabHallA:1999epl,JeffersonLabHallA:2005wfu,JeffersonLabHallA:2007ati} for studying polarization transfer in elastic reactions. More recently, proposals for measuring the recoil proton spin have been put forward in the context of Deeply Virtual Compton Scattering   
\cite{BessidskaiaBylund:2022qgg}, and similar measurements   may be possible in  future at the  Electron-Ion Collider (EIC) \cite{henry}. Given these developments mostly on the experimental side, the time may be ripe to revisit the old proposal from  modern perspective.  Our work enriches the physics of transverse spin and builds  additional physics cases made possible by the ability to measure  recoil baryon polarization.

\section{T-odd fracture function}

We consider the reaction  $e(\ell)+p(p)\to e'(\ell')+B(p')+X$ in unpolarized DIS where $B$ is the spin-$\frac{1}{2}$ leading baryon (i.e., the one with the largest longitudinal momentum fraction) such as the proton $p$,  neutron $n$ or  Lambda $\Lambda$ detected in the target fragmentation region.  We assume that $p'$ is almost collinear to the incoming proton beam  such that the `momentum transfer' $-t\equiv -(p-p')^2$ is negligible compared to the photon virtuality $Q^2=-q^2=-(\ell-\ell')^2$. We further assume that  the baryon $B$  is transversely polarized with the spin vector $s'^\mu \approx \delta^\mu_\alpha s'^\alpha_\perp$ ($\alpha=1,2$,  normalized by the baryon mass $|s'_\perp|=m_B$). Such `spontaneous' polarization is well known in the context of $\Lambda$-production in unpolarized proton-proton and proton-nucleus collisions \cite{Bunce:1976yb,Lundberg:1989hw}, but the phenomenon is more general and occurs for other  baryons and  in other processes. 

Let us introduce the relevant hadronic tensor 
\beq
W^{\mu\nu}(s')=\frac{1}{4\pi}\sum_X\int d^4y e^{-iq\cdot y}\langle p|J^\mu(0)|B(p's')X\rangle\langle B(p's')X|J^\nu(y)|p\rangle,
\eeq
where $J^\mu=\sum_f e_f \bar{\psi}_f\gamma^\mu \psi_f$ is the electromagnetic current.  The differential cross section is given by
\beq
\frac{d\sigma}{dx_BdQ^2d\phi_{\ell'} dx_L dt d\phi_{p'}}= \frac{\alpha_{em}^2y^2}{32\pi^3 Q^6}  L^{\mu\nu}W_{\mu\nu}, \label{dsigma}
\eeq
where $L^{\mu\nu}=2(\ell^\mu \ell'^\nu + \ell^\nu \ell'^\mu-g^{\mu\nu}\ell \cdot \ell')$ is the unpolarized lepton tensor.  $y=\frac{q\cdot p}{\ell\cdot p}$ and $x_B=\frac{Q^2}{2p\cdot q}$ are the standard DIS variables and  we define the longitudinal momentum fraction as $x_L=p'^+/p^+$.

The general parametrization of $W^{\mu\nu}$ consistent with the symmetries of  QCD reads 
\beq
W^{\mu\nu}=F_1(x_B,x_L,t)\left(-g^{\mu\nu}+\frac{q^\mu q^\nu}{q^2}\right) + \frac{F_2(x_B,x_L,t)}{p\cdot q} \left(p^\mu-\frac{p\cdot q}{q^2}q^\mu \right)\left(p^\nu-\frac{p\cdot q}{q^2}q^\nu \right)
\nn
+\frac{F_{\rm odd}(x_B,x_L,t)}{(p\cdot q)^2} \left\{\left(p^\mu-\frac{p\cdot q}{q^2}q^\mu \right) \epsilon^{\nu p q s'} + \left(p^\nu-\frac{p\cdot q}{q^2}q^\nu \right) \epsilon^{\mu p q s'}\right\}+\cdots. \label{wmu}
\eeq
 $p'^\mu$  does not constitute an independent vector since it is nearly parallel to $p^\mu$ by assumption.  
The spin-independent fracture functions $F_{1,2}(x_B,x_L,t)$ \cite{Trentadue:1993ka}  reduce to the usual DIS structure functions $F_{1,2}(x_B)$ upon integration over the phase space of $B$ and summation over all baryon species $B$.  $F_{\rm odd}$ is an example of the `T-odd fracture function' (TOFF) \cite{Teryaev:2002wf} which is the main focus of this work. 

The  $q^2$ poles in $W^{\mu\nu}$ should cancel in the real photon limit  $q^2 \to 0$, as there are no strongly interacting massless particles. This cancellation  will result in a Callan-Gross type  relation, completely similar to that for the DIS structure functions. Recall, that it is owing to the transverse polarization of virtual photon and is valid in the limits of both small (where photon is almost real and therefore almost transverse) and large (when only transverse photons interact with 
quarks with the mass much smaller than their virtuality) $q^2$. For TOFF $F_{\rm odd}$, there is no 
possibility of such a cancellation, and we should expect that it goes to zero in the real photon limit:
\beq
F_{\rm odd}(x_B,x_L,t)|_{x_B \to 0} \to 0. \label{CG} 
\eeq
As function $F_{2}(x_B,x_L,t)$ exhibits the same behavior, due to Callan-Gross relation, the ratio 
$F_{\rm odd}(x_B,x_L,t)/F_{2}(x_B,x_L,t)$ should remain finite in that limit.

Unlike previous applications of TOFF associated with initial proton polarizations \cite{Teryaev:2002wf,Anselmino:2011ss,Chen:2021vby,Chen:2023wsi}, here $T_{\rm odd}$ accompanies the spin vector $s'^\mu$ of the outgoing baryon.\footnote{Fracture functions tied to final state polarization have been previously considered in  \cite{Anselmino:2011ss,Xi:2025feb}, but only the  leading-twist, T-even ones.}  Inspection of the tensor structure shows that $F_{\rm odd}$  arises from  the interference between the transversely ($T$) and longitudinally ($L$) polarized virtual photon.  Plugging (\ref{wmu}) into (\ref{dsigma}), we find 
\beq
\frac{d\sigma^{LT}}{dx_BdQ^2d\phi_{\ell'} dx_L dt d\phi_{p'}}= \frac{\alpha_{em}^2(2-y)}{4\pi^3s_{ep}Q^6}\epsilon^{p q \ell's'}F_{\rm odd}(x_B,x_L,t). \label{ssa}
\eeq
In a frame in which $p^\mu$ and $q^\mu$ are collinear, (\ref{ssa}) represents an azimuthal asymmetry   between the scattered lepton momentum and the transverse spin of the  outgoing baryon  
\beq
\frac{d\sigma}{d\phi_{\ell'}} \sim \sin (\phi_{\ell'}-\phi_{s'}). \label{phil}
\eeq
This type of correlation was predicted by Christ and Lee  \cite{Christ:1966zz} for the special case of $e+p\to e'+N^*$. More precisely, they considered the subsequent decay process  $N^*\to N+\pi$ since this is the only way to access the spin information of $N^*$. Such a decay produces  an absorptive part which interferes with the non-resonance continuum $e+p\to e'+N+\pi$,  leading to the asymmetry (\ref{phil}).\footnote{As also observed in \cite{Christ:1966zz}, this type of background can be largely eliminated  by integrating over the azimuthal angle $\phi_{p'}$ as in (\ref{phi}). 
}  As we shall see in the following, in inelastic events with particle production, there are other types of final state interactions that generate absorptive parts and  contribute to $F_{\rm odd}$   even when $B$ is the ground state nucleon. 

 In the next sections, we discuss two limits in which we can gain some analytical insights into the nonperturbative function $F_{\rm odd}$. These are the high-$Q^2$ limit (Section III) and the Regge limit (Section IV). Originally the asymmetry was considered in the laboratory (lepton-proton collinear) frame \cite{Christ:1966zz}. Theoretically, it is cleaner to work in frames in which the virtual photon and the proton are collinear, and this will be assumed throughout.

\begin{figure}[t]
\vspace{-25mm}
    \centering
    \hspace{10mm}\includegraphics[width=1\linewidth]{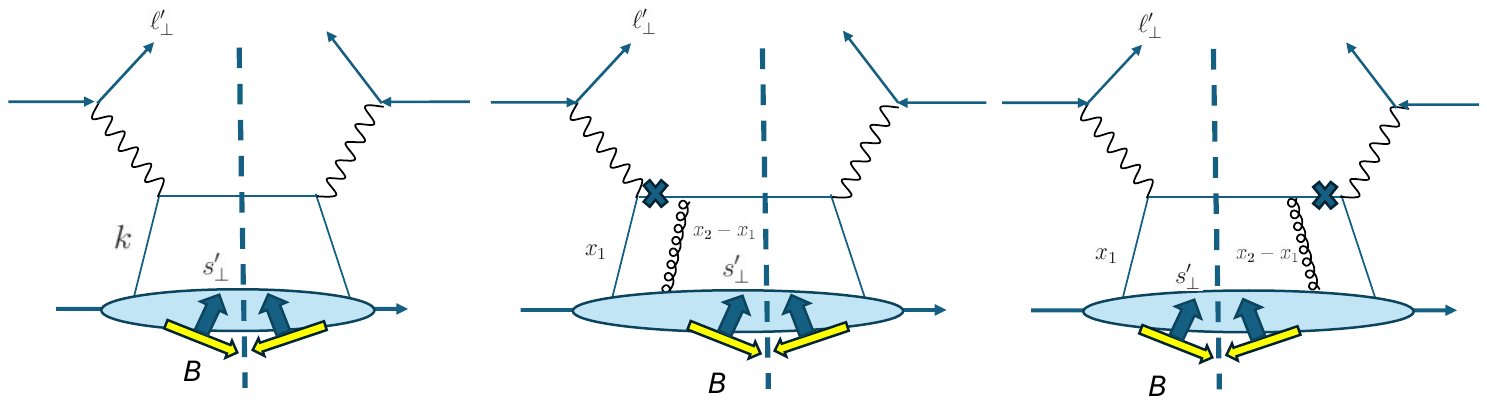}
    \vspace{-60mm}
    \caption{Lowest order diagrams for the process  $e+p\to e'+B^\uparrow+X$. The cross on the quark line represents the on-shell imaginary part. }
    \label{frac}
\end{figure}

\section{Twist-three fracture function}

In DIS at high-$Q^2$, one can apply perturbative QCD approaches  for describing an asymmetry of the form (\ref{phil}). The relevant diagrams are shown in Fig.~\ref{frac}. A cross in the last two diagrams denotes the imaginary (pole) part of the quark propagator. It is well known that such a phase is necessary for  generating `ordinary' SSAs \cite{Efremov:1981sh,Efremov:1984ip,Qiu:1991pp,Qiu:1991wg}. Importantly, in the context of fracture functions,  the left diagram also contributes even at the tree level. This was first observed in \cite{Chen:2023wsi} in the case of initial state polarization. The discussion for final state polarization turns out to be quite analogous.  Following the standard procedure \cite{Efremov:1983eb,Ratcliffe:1985mp,Qiu:1998ia,Eguchi:2006mc,Xing:2019ovj,Benic:2019zvg}, one can decompose the hadronic tensor into three parts 
\beq
W^{\mu\nu}=W_1^{\mu\nu}+W_2^{\mu\nu}+W_3^{\mu\nu}.
\eeq
$W_1$ is obtained by taking the collinear limit $k^\mu = xp^\mu$ of the left diagram in Fig.~\ref{frac} and reads 
\beq
W^{\mu\nu}_1= \frac{1}{4\pi} \sum_f e_f^2  \int dx  H_{ji}^{\mu\nu}(xp)\int \frac{ d\mu}{2\pi}e^{i\mu x} \sum_X\langle p|\bar{\psi}^f_j(0)W_{0\infty}|B(p's')X\rangle\langle B(p's')X| W_{\infty \mu}\psi^f_i(\mu n)|p\rangle,
\label{w1}
\eeq
where $e_f$ is the electromagnetic charge (in units of $|e|$) of  quark flavor $f$, and  $i,j$ are Dirac indices. $n^\mu=\delta^\mu_-/p^+$ is the conjugate vector to $p^\mu$ such that $p\cdot n=1$, and  $W$'s  denote Wilson lines extending to  future infinity along the direction $n^\mu$.   The hard factor is 
\beq
H^{\mu\nu}(k)=2\pi\delta((k+q)^2)\gamma^\mu(\Slash k+\Slash q)\gamma^\nu.
\eeq
The spin-dependent part of the  collinear twist-three fracture function can be parametrized as   
\beq
\int \frac{ d\mu}{2\pi}e^{i\mu x}\sum_X \langle p|\bar{\psi}_j(0)|B(p's')X\rangle\langle B(p's')X| \psi_i(\mu n)|p\rangle = (\gamma_5\Slash s')_{ij} g_{TR}(x,x_L,t)+\epsilon^{pn \alpha\beta}(\gamma_\alpha)_{ij} s'_\beta u_{TR}(x,x_L,t). \label{gt}
\eeq
$g_{TR}$ ($R$ for `recoil') is an analog of the familiar $g_T(x)$ distribution for the transversely polarized nucleon. On the other hand, the appearance of the function $u_{TR}$ may seem surprising at first.  (The corresponding function in the case of initial state polarization is denoted by $u_T$  in  \cite{Chen:2023wsi}.) Naively, $u_{TR}$ violates T-invariance and is nonzero only because of final state interactions encoded in the Wilson line $W$. 
Substituting (\ref{gt}) into (\ref{w1}), we obtain
\beq
W_1^{\mu\nu}= \sum_f e_f^2 \frac{1}{(p\cdot q)^2} \left((x_Bp+q)^\mu \epsilon^{\nu pq s'}+(x_Bp+q)^\nu \epsilon^{\mu pq s'}\right)u^f_{TR}(x_B,x_L,t), \label{w1fin}
\eeq
where we wrote $\epsilon^{\nu p n s'}=\frac{1}{p\cdot q}\epsilon^{\nu p q s'}$ assuming that we work in frames where the virtual photon and the incoming proton are collinear. The $g_{TR}$ term is neglected in (\ref{w1fin}) since it is purely imaginary. At higher order in perturbation theory, the hard factor $H^{\mu\nu}$ acquires an imaginary part, and the  $g_{TR}$ term should also contribute, cf.,    \cite{Benic:2019zvg,Benic:2021gya,Benic:2024fvk}.

The $W_2$ term comes from the left diagram when it is expanded around the collinear limit $k^\mu = xp^\mu+k^\mu_\perp$, as well as part of the last two diagrams. A factor of $k_\perp^\mu$ is converted into a derivative acting on the quark field, giving 
\beq
W_2^{\mu\nu}&=&
\frac{i}{4\pi} \sum_f e_f^2\int dx \left. \pi \frac{\partial}{\partial k_{\perp}^\alpha}H^{\mu\nu}_{ji}(k)\right|_{k=xp} \int \frac{d\lambda }{2\pi}e^{i\lambda x}  \nn
&&\qquad \qquad \times 
 \sum_X \langle p|\bar{\psi}^f_j(0)W_{0\infty}|B(p's')X\rangle\langle B(p's')X|D_\perp^\alpha(\infty)W_{\infty \lambda} \psi^f_i(\lambda n)|p\rangle, \label{w}
\eeq
where the index $\alpha=1,2$ is transverse. The covariant derivative $D_\perp^\alpha(\infty)$ may be replaced by the ordinary derivative $\partial_\perp^\alpha$ assuming that the gauge field vanishes at infinity. The matrix element can be parametrized as 
\beq
i\int \frac{d\lambda }{2\pi}e^{i\lambda x}  
 \sum_X \langle p|\bar{\psi}_j(0)|B(p's')X\rangle\langle B(p's')X|D_\perp^\alpha W \psi_i(\lambda n)|p\rangle   = (\Slash p)_{ij}\epsilon^{\alpha pn s'}u_{\partial TR}(x,x_L,t)-(\gamma_5\Slash p)_{ij}s'^\alpha \tilde{g}_R(x,x_L,t), \label{gd}
\eeq
Again, the $u_{\partial TR}$ term is naively forbidden by T-invariance, while the function $\tilde{g}_R$ has an analog in the context of ordinary SSA (called $\tilde{g}$ in  \cite{Eguchi:2006mc}). 
The derivative acting on the delta function can be handled by writing 
\beq
\left.\frac{\partial}{\partial k_{\perp}^\alpha} \delta(k^2+2k\cdot q+q^2)\right|_{k=xp}
=\frac{2q_{\perp\alpha}}{(2p\cdot q)^2} \frac{d}{dx}\delta(x-x_B),
\eeq
but this vanishes in photon-proton collinear frame $q_\perp=0$. Only the first term in (\ref{gd})  contributes to the symmetric part and we find 
\beq
W_2^{\mu\nu}=\frac{\pi}{(p\cdot q)^2}\sum_f e_f^2 u_{\partial TR}^f(x_B) \left(p^\mu \epsilon^{\nu p q s'}+p^\nu \epsilon^{\mu p q s'}\right) .\label{w2fin}
\eeq
Finally, $W_3$ comes from the last two diagrams in Fig.~\ref{frac} picking up the pole part $\delta((k_{1,2}+q)^2)$  of the quark propagator 
\beq
W_3^{\mu\nu}&=&\frac{1}{4\pi p^+} \sum_f e_f^2\sum_X \int dx_1dx_2 \int \frac{d\lambda d\mu}{(2\pi)^2}e^{i\lambda x_1+i\mu(x_2-x_1)} \left. \pi \frac{\partial}{\partial k_{2\perp}^\alpha}{\cal H}^{\mu\nu}_{ji}(k_1,k_2)\right|_{k^\mu_i=x_ip^\mu} \label{w}\\
&&\times \Biggl[
 \langle p|\bar{\psi}^f_j(0)|B(p's')X\rangle\langle B(p's')X|gF^{\alpha+}(\mu n) \psi^f_i(\lambda n)|p\rangle -\langle p|\bar{\psi}^f_j(0)gF^{\alpha+}(\mu n)|B(p's')X\rangle\langle B(p's')X|\psi^f_i(\lambda n)|p\rangle  \Biggr] , \notag
\eeq
where 
\beq
{\cal H}^{\mu\nu}(k_1,k_2)=2\pi\delta((k_1+q)^2)\delta((k_2+q)^2)\gamma^\mu (\Slash k_2+\Slash q)\Slash p(\Slash k_1+\Slash q)\gamma^\nu.  
\label{hard}
\eeq
 The matrix element can be parametrized as 
 \beq
\frac{1}{p^+}\int \frac{d\lambda d\mu}{(2\pi)^2}e^{i\lambda x_1+i\mu(x_2-x_1)}\sum_X \langle p|\bar{\psi}_j(0)|B(p's')X\rangle\langle B(p's')X|gF^{\alpha+}(\mu n)\psi_i(\lambda n)|p\rangle \nn = (\Slash p)_{ij} \epsilon^{\alpha p n s'}G_{FR}(x_1,x_2,x_L,t)-i(\gamma_5\Slash p)_{ij} s'^\alpha \tilde{G}_{FR}(x_1,x_2,x_L,t), \label{1}
\eeq
\beq
\frac{1}{p^+}\int \frac{d\lambda d\mu}{(2\pi)^2}e^{i\lambda x_1+i\mu(x_2-x_1)}\sum_X \langle p|\bar{\psi}_j(0)gF^{\alpha+}(\mu n)|B(p's')X\rangle\langle B(p's')X|\psi_i(\lambda n)|p\rangle \nn = (\Slash p)_{ij} \epsilon^{\alpha p n s'}G_{FR}^*(x_2,x_1,x_L,t)+i(\gamma_5\Slash p)_{ij} s'^\alpha \tilde{G}_{FR}^*(x_2,x_1,x_L,t). \label{2}
\eeq
(\ref{2}) follows from (\ref{1}) by hermiticity. In contrast to (\ref{gt}) and (\ref{gd}), neither   $G_{FR}$ nor $\tilde{G}_{TR}$ is forbidden by  the naive application of T-invariance.  These are the counterparts of the Efremov-Teryaev-Qiu-Sterman (ETQS) functions $ \langle ps|\bar{\psi}gF^{+\alpha}\psi|ps\rangle$ \cite{Efremov:1981sh,Efremov:1984ip,Qiu:1991pp,Qiu:1991wg} relevant to SSAs with initial polarization. The difference is that, due to final state interactions,  T-invariance does not imply that   $G_{FR}$ and $\tilde{G}_{FR}$ are real functions.    
Yet, in (\ref{w}) a particular  final state interaction and the resulting phase have been explicitly separated out in the hard part ${\cal H}$. By focusing on this `soft gluon pole'  contribution, we treat $G_{FR}$ and $\tilde{G}_{FR}$ as real. Generalization to  complex $G_{FR},\tilde{G}_{FR}$ can also be done following   \cite{Chen:2023wsi}.

Substituting (\ref{1}), (\ref{2}) into  (\ref{w}), we find   
\beq
W^{\mu\nu}=\frac{\pi}{4p\cdot q}\sum_fe_f^2\int dx_2 \frac{\partial}{\partial k^\alpha_{2\perp}} \Biggl[\delta((k_2+q)^2) \biggl\{ {\rm Tr}[\gamma^\mu(\Slash k_2+\Slash q)\Slash p(x_1\Slash p+\Slash q)\gamma^\nu\Slash p](G_{FR}^f(x_B,x_2)-G_{FR}^f(x_2,x_B)) \epsilon^{\alpha p n s'}\nn 
-is'^\alpha\,{\rm Tr}[\gamma^\mu(\Slash k_2+\Slash q)\Slash p(x_1\Slash p+\Slash q)\gamma^\nu\gamma_5\Slash p](\tilde{G}_{FR}^f(x_B,x_2)+\tilde{G}_{FR}^f(x_2,x_B)) \biggr\}\Biggr]_{k_2=x_2p}. \label{new}
\eeq
It is easy to check that the $G_{FR}$ term vanishes in photon-proton collinear frames.\footnote{In the laboratory frame where $\ell'_\perp = -q_\perp$, the first term in (\ref{new})  leads to a nonzero contribution proportional to 
\beq
\epsilon^{q p n s'}\left(-g^{\mu\nu}+\frac{p^\mu q^\nu + p^\nu q^\mu}{p\cdot q}-\frac{q^2}{(p\cdot q)^2}p^\mu p^\nu\right)\left.\frac{d}{dx}(G_{FR}(x,x_B)-G_{FR}(x_B,x))\right|_{x=x_B}.
\label{drop}
\eeq
Note that the tensor structure is that of the transversely polarized virtual photon rather than the transverse-longitudinal interference effect that we identified in the previous section. (\ref{drop}) contains  the same angular correlation $\sin(\phi_{\ell'}-\phi_{s'})$ and contaminates the frame-independent effect coming from the second term of (\ref{new}).  } 
As for the $\tilde{G}_{FR}$ term, the only nonvanishing contribution (after neglecting irrelevant antisymmetric terms $\epsilon^{\mu\nu\cdots}$) arises when the derivative acts on $\Slash k_2$ inside the trace, giving 
\beq
W_3^{\mu\nu}=\frac{2\pi}{(p\cdot q)^2} \sum_fe_f^2\tilde{G}_{FR}^f(x_B,x_B) p^\mu \epsilon^{\nu p q s'}+\cdots. \label{w3fin}
\eeq 
This can be made symmetric in $\mu$ and $\nu$  by using the Schouten identity
\beq
p^\mu\epsilon^{\nu p q s'} =\frac{1}{2}(p^\mu \epsilon^{\nu p q s'}+p^\nu \epsilon^{\mu p q s'}+p\cdot q \epsilon^{\mu\nu ps'}),
\eeq
and dropping the antisymmetric term. 

The three contributions (\ref{w1fin}), (\ref{w2fin}), (\ref{w3fin}) can be combined using the  relation \cite{Chen:2023wsi}
\beq
xu_{TR}(x)= u_{\partial TR}(x) +\pi\tilde{G}_{FR}(x,x), \label{eom}
\eeq
that follows from the Dirac equation 
\beq
0=\gamma^\alpha_\perp \gamma^+\Slash D\psi =(\gamma^\alpha_\perp -i\epsilon^{\alpha\beta}\gamma_\beta\gamma_5)D_-\psi  -\gamma^+D^\alpha_\perp \psi + i\epsilon^{\alpha\beta}\gamma^+\gamma_5D_\beta \psi,
\eeq
together with the formula\footnote{One can write $\theta(\tau-\lambda)=\frac{1}{2}(1+\epsilon(\tau-\lambda))$ where $\epsilon(\tau)$ is the sign function. If ordinary SSA where one  deals with  the matrix element $\langle ps|\bar{\psi}gF^{+\alpha}\psi|ps\rangle$, the `$\frac{1}{2}$' term can be neglected \cite{Eguchi:2006mc}. However, in the present problem it results in  the $\tilde{G}_{FR}(x,x)$ term in (\ref{eom}).  } 
\beq
W_{\infty \lambda}D_\perp^\alpha(\lambda n)\psi(\lambda n )=D_\perp^\alpha(\infty)W_{\infty \lambda}\psi(\lambda   n)+\frac{ig}{p^+}\int d\tau \theta(\tau-\lambda) W_{\infty \tau}F^{\alpha +}(\tau  n)W_{\tau\lambda}\psi(\lambda n).
\eeq
The result is 
\beq
W^{\mu\nu}= \sum_f e_f^2 \frac{1}{(p\cdot q)^2} \left((2x_Bp+q)^\mu \epsilon^{\nu pq s'}+(2x_Bp+q)^\nu \epsilon^{\mu pq s'}\right)u^f_{TR}(x_B,x_L,t), \label{wfin}
\eeq
or equivalently (cf. (\ref{CG})), 
\beq
F_{\rm odd}(x_B,x_L,t)=2\sum_f e_f^2 x_B u^f_{TR}(x_B,x_L,t).
\eeq
Only the sum has the expected  QED gauge invariant tensor structure (\ref{wmu}). Combining with the lepton tensor, we find 
\beq
\frac{d\sigma}{dx_BdQ^2d\phi_{\ell'} dx_L dt d\phi_{p'}}=\sum_f e_f^2\frac{\alpha_{em}^2(2-y)}{4\pi^3 s_{ep}Q^6} x_Bu_{TR}^f(x_B,x_L,t) \epsilon^{pq\ell' s'}.  \label{inco}
\eeq
In practice, we may integrate over $x_L$ and  $t$, assuming that the integral is dominated by the nonperturbative region $-t\lesssim 1$ GeV$^2$. Due to parity, the baryon will be  polarized perpendicularly to the production plane $\phi_{s'}=\phi_{p'}+\frac{\pi}{2}$ (or $\phi_{s'}=\phi_{p'}-\frac{\pi}{2}$). We may further integrate over $\phi_{p'}$ keeping  the relative angle $\phi=\phi_{\ell'}-\phi_{s'}$ fixed 
\beq
\frac{d\sigma}{dx_BdQ^2d\phi}= \sin \phi \frac{\alpha_{em}^2 (2-y)\sqrt{1-y}m_B }{2\pi^2ys_{ep} Q^5}\sum_f e_f^2 x_Bu_{TR}^f(x_B) . \label{phi}
\eeq

(\ref{wfin}) is formally identical, after the replacement $u_{TR}\to u_T$, to the corresponding  result  for  $W^{\mu\nu}\sim \langle ps|J^\mu|hX\rangle\langle hX|J^\nu|ps\rangle$ \cite{Chen:2023wsi} where the authors considered initial proton polarization  and arbitrary hadron $h$ detection in the target fragmentation region. However, $u_T$ and $u_{TR}$ are different functions even when $h=B$. In both cases, only one fracture function $u_T$ or $u_{TR}$ is responsible for the asymmetry, in contrast to SSAs in the current fragmentation region  where both  the ETQS functions $ \langle ps|\bar{\psi}gF^{+\alpha}\psi|ps\rangle$ \cite{Efremov:1981sh,Efremov:1984ip,Qiu:1991pp,Qiu:1991wg}, and the twist-three fragmentation functions $\langle 0|\bar{\psi}F^{+\alpha}|hX\rangle\langle hX|\psi|0\rangle$ \cite{Koike:2001zw,Boer:2003cm}  contribute. In a sense, the fracture function is a hybrid of the ETQS and fragmentation functions. 
Another notable feature is that,  since no high-$p_T$ hadron is required in the final state, (\ref{wfin}) hence also the asymmetry is zeroth order in $\alpha_s$. This means that the corrections to (\ref{wfin})  coming from the $g_{TR}$ term in (\ref{gt})  mentioned below (\ref{w1fin})  will be suppressed by $\alpha_s^2$ relative to the leading term.

\section{Pomeron-odderon interference}

Our second example is diffractive DIS in the Regge limit $p\cdot q\gg Q^2$, in which the proton scatters elastically and loses a small fraction of energy $x_L\approx 1$.
$h=p$ and $X$ are separated by a rapidity gap $Y_{\rm gap}=\ln \frac{1}{1-x_L}$. 
In this setup, we show that an SSA of the form (\ref{phil}) arises from the interference between the Pomeron and the odderon.

Again we work in a photon-proton collinear frame. At high energy, this is sometimes referred to as the dipole frame because the leading contribution comes from the process in which the virtual photon splits into a quark-antiquark pair (`color dipole') $q=p_q+p_{\bar{q}}$ with the momentum partition $p_q^-=zq^-$ and $p_{\bar{q}}^-=(1-z)q^-\equiv \bar{z}q^-$. The pair then scatters off the proton with a color-singlet exchange in the $t$-channel. The cross section takes the general form  
\beq
\frac{d\sigma}{dx_BdQ^2 d\phi_{\ell'} dz d^2p_{q\perp} dtd\phi_{p'}}=\frac{\alpha_{em}}{2\pi^2 x_B Q^2}\sum_f e_f^2\Biggl[\frac{1+(1-y)^2}{2} \frac{d\sigma^T}{dzd^2p_{q\perp} dtd\phi_{p'}} + (1-y)\frac{d\sigma^L}{dzd^2p_{q\perp} dt d\phi_{p'}} \nn +\frac{y(2-y)}{2} \frac{d\sigma^{LT}}{dzd^2p_{q\perp} dtd\phi_{p'}}+\cdots\Biggr]. \label{general}
\eeq
 In the near-forward kinematics $t\approx 0$, the quark and the antiquark have back-to-back transverse momenta $k_\perp$ and $-k_\perp$ in this frame. 
The transverse-longitudinal interference term explicitly reads  ($\mu^2\equiv z\bar{z}Q^2$) 
\beq
 \frac{d\sigma^{LT}}{dzd^2p_{q\perp} dt d\phi_{p'}} &=&-\frac{\alpha_{em}}{4N_c}\frac{\ell'^i}{(q^-)^2}\int d^2k_\perp d^2k'_\perp \frac{|T|^2}{((p_{q\perp}-k_\perp)^2+\mu^2)((p_{q\perp}-k'_\perp)^2+\mu^2)}\nn
&& \times  \sum_{\rm spin}\Biggl[\bar{v}(p_{\bar{q}})\gamma^-u(p_q) 
\bar{u}(p_q)((\bar{z}-z)(p_q^i-k^i)\gamma^- + i\epsilon^{ij}(p_{qj}-k_j)\gamma^-\gamma_5)v(p_{\bar{q}}) \nn 
 && \qquad +\bar{v}(p_{\bar{q}})((\bar{z}-z)(p_q^i-k'^i)\gamma^- + i\epsilon^{ij}(p_{qj}-k'_j)\gamma_5\gamma^-)u(p_{q}) \bar{u}(p_q)\gamma^-v(p_{\bar{q}}) \Biggr] \nn 
 &=&-\frac{4\alpha_{em} }{N_c}(\bar{z}-z)z\bar{z}\int d^2k_\perp d^2k'_\perp \frac{\ell'^i (p_q^i-k^i)|T|^2}{((p_{q\perp}-k_\perp)^2+\mu^2)((p_{q\perp}-k'_\perp)^2+\mu^2)}, \label{had}
\eeq
where  quark masses have been neglected. 
The symbol $|T|^2$  represents the squared $T$-matrix describing the near-forward scattering between  the $q\bar{q}$ pair and the proton. It is an abbreviation of \cite{Boussarie:2019vmk,Hagiwara:2020mqb}
\beq
|T|^2 &\equiv& \frac{1}{2}\sum_s \frac{1}{2m_p^2}\bar{u}(p's')\left(T(k_\perp) +\frac{\alpha_s}{2k_\perp^2}\frac{\sigma^{i+}}{p^+}k^i O(k_\perp)\right)u(ps)\bar{u}(ps)\left(T(k'_\perp) +\frac{\alpha_s}{2k'^2_\perp}\frac{\sigma^{i+}}{p^+}k'^i O(k'_\perp)\right)u(p's') \nn 
&\approx &T(k_\perp)T(k'_\perp)-\alpha_s\frac{\epsilon^{ij}}{2m_p^2}\left(\frac{k^i}{k_\perp^2} O(k_\perp)T(k'_\perp)+ \frac{k'^i}{k'^2_\perp} T(k_\perp)O(k'_\perp)\right) s'^j.  \label{tt}
\eeq
The spin-independent $C$-even part  $T(k_\perp)$ 
is the dipole $T$-matrix associated with the  `Pomeron' exchange.   
The $C$-odd part $O$ is the spin-dependent odderon \cite{Zhou:2013gsa,Boer:2015pni,Dong:2018wsp,Yao:2018vcg,Boussarie:2019vmk,Hagiwara:2020mqb,Kovchegov:2021iyc,Benic:2024fbf}. In the forward limit, it is related to the gluon Sivers function  as
\beq
O(x,k_\perp)= -xf_{1T}^{\perp g}(x,k_\perp).
\eeq

Let us integrate (\ref{had}) over $p_{q\perp}$. The $TT$ term in (\ref{tt}) drops out due to azimuthal symmetry. The second, Pomeron-odderon interference term gives, after the angular integrals over $\phi_{k}$, $\phi_{k'}$ and $\phi_{p_q}$, 
\beq
\frac{d\sigma^{LT}}{ dzdp^2_{q\perp}dtd\phi_{p'}} = \frac{\pi\alpha_{em}\alpha_s}{N_cm_p^2} \epsilon^{ij}\ell'^i s'^j (\bar{z}-z)z\bar{z}  \Biggl[ \int \frac{d^2k_\perp}{k_\perp^2} \frac{(2p_{q\perp}-k_{\perp})\cdot k_\perp O(k_\perp) }{(p_{q\perp}-k_\perp)^2+\mu^2}\int  \frac{ d^2k_\perp T(k_\perp)}{(p_{q\perp}-k_\perp)^2+\mu^2} \nn 
 - \int \frac{d^2k_\perp}{k_\perp^2} \frac{p_{q\perp}\cdot k_\perp O(k_\perp) }{(p_{q\perp}-k_\perp)^2+\mu^2}\int \frac{d^2k_\perp}{p_{q\perp}^2} \frac{  p_{q\perp}\cdot k_\perp T(k_\perp)}{(p_{q\perp}-k_\perp)^2+\mu^2} \Biggr]. \label{zz}
\eeq
We have thus found the correlation (\ref{phil}). To be as inclusive as possible, it is tempting to further integrate (\ref{zz}) over $0\le z\le 1$. This corresponds to inclusive diffraction summed over all (diffractive) final states $X$. However, the asymmetry vanishes in this case since the integrand is antisymmetric under $z\to \bar{z}$. In physical terms, the Pomeron-odderon interference is C-parity odd, and one needs to select final states without definite C-parity \cite{Brodsky:1999mz,Hagler:2002nh,Hagler:2002nf,Benic:2025okp,Mantysaari:2025mht}. One way to do so is to measure the differential cross section $d\sigma/dz$, but this requires discriminating quark jets from antiquark jets, which is difficult to perform in practice.  

Alternatively, and more practically, one can inclusively  measure a positively (or negatively) charged hadron species, say $\pi^+$, in the  produced system $X$. This is an example of semi-inclusive `diffractive' DIS (SIDDIS)  \cite{Hatta:2022lzj,Guo:2023uis,Fucilla:2023mkl,Bhattacharya:2024sck,Hatta:2024vzv,Teryaev:2001zu}. The cross section is obtained by convoluting  (\ref{zz}) with the valence quark fragmentation function (see e.g. \cite{Kotlorz:2025xso}) appearing due to the mentioned C-parity oddness:    
\beq
\left.\frac{d\sigma^{LT}}{dz_FdP^2_{h\perp}dtd\phi_{p'}} =\sum_f e_f^2\int_{z_F}^1 \frac{dz}{z}\frac{z_F^2}{z^2}\left(D_{f\to \pi^+}\left(\frac{z_F}{z}\right)- D_{\bar{f}\to \pi^+}\left(\frac{z_F}{z}\right)\right) \frac{d\sigma^{LT}}{dzdp^2_{q\perp}dt d\phi_{p'}}\right|_{p_{q\perp}=\frac{z}{z_f}P_{h\perp}}. \label{siddis}
\eeq
This  is nonvanishing  even if isospin invariance is assumed $D_{u\to \pi^+}(z)=D_{\bar{d}\to \pi^+}(z)$  because of the charge factor $e^2_u\neq e^2_d$. One may sum over all positively charged hadrons to increase the rates.  

\begin{figure}[t]
        \begin{overpic}[width=0.5\textwidth]{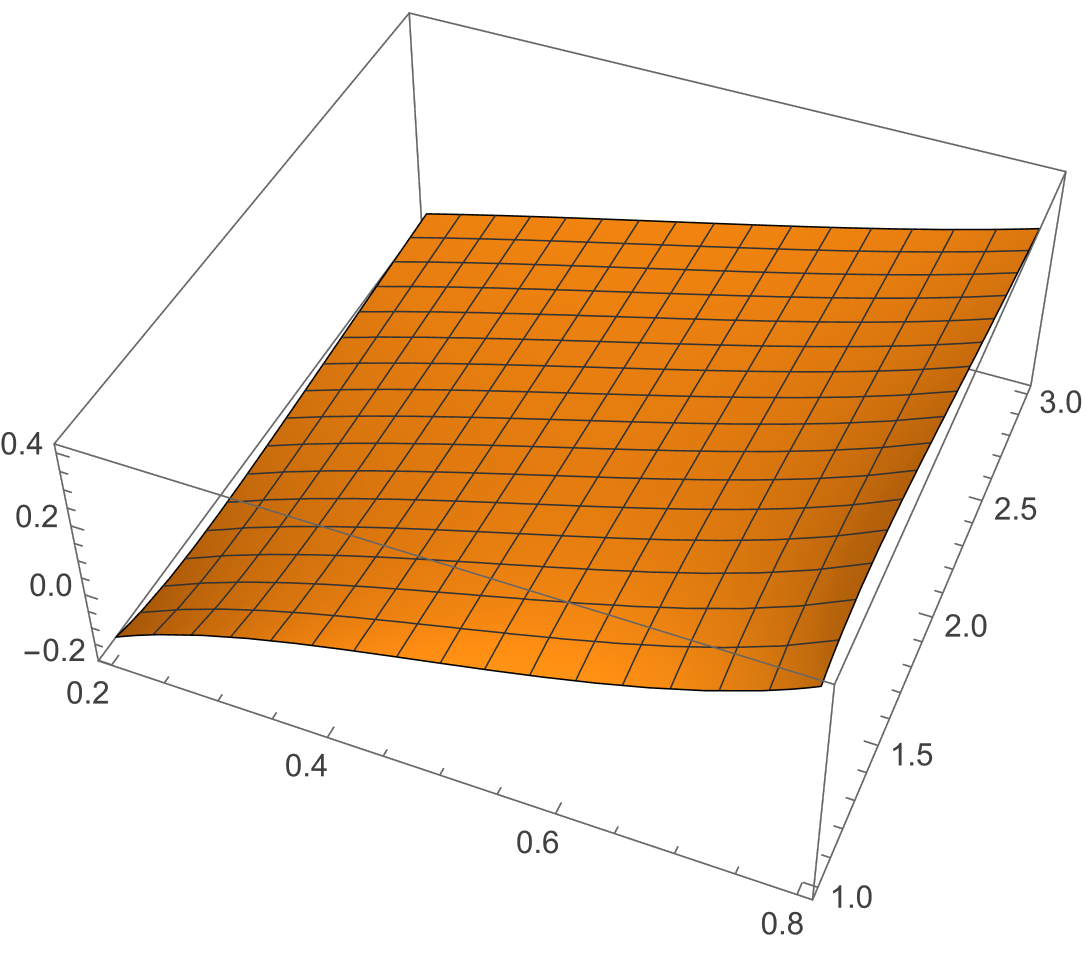}
            \put(-8,35){\Large $\frac{A_N}{C}$}
            \put(92,20){\rotatebox{70}{\large $P_{h\perp}^2\ (\mathrm{GeV}^2)$}}
            \put(35,8){\rotatebox{-8}{ \Large $z_f$}}
        \end{overpic}
        \caption{SSA $A_N$ (divided by the constant $C$) as a function of $z_f$ 
        and $P_{h\perp}^2$. 
        } \label{pic}
  \end{figure}

At the moment, almost nothing is known about the magnitude of the spin-dependent odderon. Let us nevertheless attempt at a very crude estimate of the asymmetry normalized by the first two term of (\ref{general}).
\beq
\left.\frac{d\sigma^{T,L}}{dz_FdP^2_{h\perp}dtd\phi_{p'}} =\sum_f e_f^2\int_{z_F}^1 \frac{dz}{z}\frac{z_F^2}{z^2}\left(D_{f\to \pi^+}\left(\frac{z_F}{z}\right)+ D_{\bar{f}\to \pi^+}\left(\frac{z_F}{z}\right)\right) \frac{d\sigma^{T,L}}{dzdp^2_{q\perp}dt d\phi_{p'}}\right|_{p_{q\perp}=\frac{z}{z_f}P_{h\perp}}. \label{siddis2}
\eeq
with 
\beq
\frac{d\sigma^T}{dzd^2p_{q\perp} dtd\phi_{p'}} = \frac{\alpha_{em}}{N_c} (z^2+\bar{z}^2) p_{q\perp}^2 \left(\int d^2k_\perp \frac{1-\frac{p_{q\perp}\cdot k_\perp}{p_{q\perp}^2}}{(p_{q\perp}-k_\perp)^2+\mu^2}T(k_\perp)\right)^2,
\eeq
and 
\beq
\frac{d\sigma^L}{dzd^2p_{q\perp} dtd\phi_{p'}} = \frac{8\alpha_{em}}{N_c} z^2\bar{z}^2Q^2 \left(\int d^2k_\perp \frac{ T(k_\perp)}{(p_{q\perp}-k_\perp)^2+\mu^2}\right)^2.
\eeq
We employ the following Gaussian  models 
\beq
T(k_\perp)=\frac{N_cS_\perp}{(2\pi)^2}\left(\delta^{(2)}(k_\perp) - \frac{1}{\pi\Lambda^2}e^{-k_\perp^2/\Lambda^2}\right),
\eeq
\beq
\alpha_s O(k_\perp) = C\frac{N_cS_\perp}{(2\pi)^2}\frac{k_\perp^2}{\Lambda_o^2}\left(2-\frac{k_\perp^2}{\Lambda_o^2}\right)e^{-k_\perp^2/\Lambda_o^2},
\eeq
where  $S_\perp$ is the proton transverse area and  $C$ is a dimensionless free parameter presumably much smaller than unity $|C|\ll 1$. 
The property 
$\int d^2k_\perp T(k_\perp)=0$ is called color transparency, while the constraint $\int d^2k_\perp O(k_\perp)=0$ follows from \cite{Zhou:2013gsa}. 
Let us fix $Q^2=2$ GeV$^2$, $y=0.5$,  $\Lambda=\Lambda_o=0.4$ GeV. The factor $N_cS_\perp$ drops out in the asymmetry ratio.  As for the pion fragmentation functions, we use the analytic parametrization $D_{u\to \pi^+}(z)=0.546 z^{-1.47}(1-z)^{1.02}$ at the scale $\mu_R^2=Q^2=2$ GeV$^2$   \cite{Kniehl:2000fe} and  neglect unfavored fragmentation channels $\bar{u}\to \pi^+,d\to \pi^+$.  In Fig.~\ref{pic}, we plot the resulting asymmetry $A_N=2\langle \sin\phi\rangle$ divided by  $C$. In order for $A_N$ to be measurable, say  $|A_N|\gg  0.1$\%, $|C|$ needs to be at least $0.01$ or larger.

The present derivation is essentially unchanged if the incoming proton is transversely polarized, as in ordinary SSA. Namely, if we  sum over $s'$ in (\ref{tt})   keeping $s^\mu=\delta^\mu_\alpha s^\alpha_\perp$ fixed  and still assuming $p^\mu \approx p'^\mu$, we obtain the same formula (\ref{zz}) (with the trivial replacement $s'\to s$) featuring the exact same function $O$. To our knowledge, this is also a new result (cf. \cite{Hatta:2024vzv}).  Of course,  it is much easier to prepare a polarized proton beam than to measure the final state proton spin. Yet, the SSA with respect to the final proton spin  is conceptually new (largely forgotten after the original work  \cite{Christ:1966zz}) and  complementary. It  can be measured  in principle even if polarization is not available (such as at HERA), or if preparing transversely polarized protons is technically challenging as in fixed target experiments. If the initial state polarization is available, the comparison of the respective asymmetry with 
the final state polarization can be an  important test of these fine effects.

\section{Conclusions}

Motivated by the prospect of measuring the polarization of  final state baryons in future DIS experiments \cite{BessidskaiaBylund:2022qgg,henry}, we have discussed the SSA of the lepton azimuthal  angle with respect to the transverse polarization of the leading baryon. In doing so, we have rediscovered the work by  Christ and Lee 60 years ago, and generalized it to a new class of SSA observables using the modern language of the nucleon structure studies. In the collinear approach at high-$Q^2$, the novel  theoretical tool is the T-odd twist-three fracture functions. In the sense described below (\ref{phi}), these are a hybrid of the twist-three distribution and fragmentation functions. Their QCD properties such as the evolution equation are yet to be explored.   In the $k_\perp$-dependent approach at high energy, the asymmetry in question can be generated by  the spin-dependent odderon. It turns out that there is no essential difference between SSAs with respect to the initial and final state polarizations. (\ref{zz}) with $s'\to s$ describes a  `usual' SSA with initial polarization, and this result is  new even in that context.   Previously, the exclusive production of  C-even mesons such as $\pi^0$ and $\chi_{c}$ in DIS has been proposed as a signature of the spin-dependent odderon  \cite{Boussarie:2019vmk,Benic:2024fbf}. In these processes, the cross section involves the square of the odderon amplitude. We expect that the present observable enjoys  a larger cross section because the Pomeron-odderon interference is linear in the odderon amplitude. 

Experimentally, final state polarization has been measured at Jefferson Lab in elastic scattering events \cite{JeffersonLabHallA:1999epl,JeffersonLabHallA:2005wfu,JeffersonLabHallA:2007ati} by observing the analyzing power of the secondary scattering of the recoil proton in a  polarimeter. Efforts are  underway to extend these measurements to DIS events, also at the EIC,   although moving to higher energies presents more significant experimental challenges \cite{BessidskaiaBylund:2022qgg}. We hope our work serves as a motivation for further technical developments in this direction.

\section*{Acknowledgments}

We are grateful to Henry Klest for motivating us to explore the physics of final state baryon polarization. 
Y.~H. is supported by the U.S. Department
of Energy under Contract No. DE-SC0012704, by LDRD funds from Brookhaven Science Associates, and also by the framework of the Saturated Glue (SURGE) Topical Theory Collaboration. 

\bibliography{ref}

\end{document}